\documentclass[12pt,showpacs,superscriptaddress,preprintnumbers,amsmath,amssymb]{revtex4}

\usepackage{epsfig} 
\usepackage{amsmath} 
\usepackage{graphicx} 
\large 
\def\bsll{B_s \rightarrow \mu^+ \, \mu^-} 
\def\bkll{B \rightarrow K \mu^+ \, \mu^-} 
 
\def\afb{\left\langle A_{FB}\right\rangle} 
\def\gev{{\rm GeV}} 
 
\newcommand{\spp}{\vphantom{\bigg(}}

\def\f{{R}}

\def\h0t{{\tilde{h}^0}}

\def\bqll{B_s \rightarrow \mu^+ \, \mu^-}

\def\alp{{ A}_{LP}} 
 
\def\h{{\cal H}}

\def\faa{\f_{\rm A}} 
\def\fpp{\f_{\rm P}} 
\def\fps{\f_{\rm S}}

\def\lesssim{\mathrel{\hbox{\rlap{\hbox{\lower4pt\hbox{$\sim$}}}\hbox{$<$}}}} 
\def\gtrsim{\mathrel{\hbox{\rlap{\hbox{\lower4pt\hbox{$\sim$}}}\hbox{$>$}}}} 
 
\begin{document}

\title{\bf Probing extended Higgs sector through \\
rare $b \to s \mu^+ \mu^-$ transitions }

\author{Ashutosh Kumar Alok}
\email{alok@theory.tifr.res.in}
\affiliation{Tata Institute of Fundamental Research, Homi Bhabha
Road, Mumbai 400005, India}

\author{Amol Dighe}
\email{amol@theory.tifr.res.in}
\affiliation{Tata Institute of Fundamental Research, Homi Bhabha
Road, Mumbai 400005, India}
                                                           
\author{S. Uma Sankar}
\email{uma@phy.iitb.ac.in}
\affiliation{Indian Institute of Technology Bombay, Mumbai 400076,
India}

%\date{\today} 

\begin{abstract} 
We study the constraints on the contribution of new physics in the 
form of scalar/pseudoscalar operators to  
the average forward backward asymmetry $\left\langle 
A_{FB}\right\rangle$ of muons in $B \rightarrow K \mu^+ \mu^-$ and 
the longitudinal polarization asymmetry $A_{LP}$ of muons in  
$B_s \to \mu^+ \mu^-$. We find 
that the maximum possible value of $\left\langle 
A_{FB}\right\rangle$ allowed by the present upper bound on  
 $B(B_s \rightarrow \mu^+ \mu^-)$ is about 
$1\%$ at $95\%$ C.L. and hence will be very difficult to measure.  
On the other hand, the present bound on 
$B(B_s \rightarrow \mu^+ \mu^-)$ fails to put 
any constraints on $A_{LP}$, which can be as high as $100\%$ 
even if $B(\bsll)$ 
is close to its standard model prediction. 
The measurement of $A_{LP}$ 
will be a direct evidence for an extended Higgs sector, 
and combined with the branching ratio 
$B(B_s \to \mu^+ \mu^-)$ it can even separate the new physics  
scalar and pseudoscalar contributions. 
\end{abstract} 

\preprint{TIFR/TH/08-15}

\pacs{13.20.He, 12.60.-i}

\maketitle

\newpage 
%%%%%%%%%%%%%%%%%%%%%%%%%%%%%%%%%%%%%%%%%%%%%%%%%%%%%%%%%%%%% 
\section{Introduction} 

The quark level flavor changing neutral interaction 
$b \rightarrow s \mu^+ \mu^-$ is forbidden at the 
tree level in the standard model (SM) and can occur only at 
the one-loop level. Therefore it can  
serve as an important probe to test 
SM at loop level and also constrain many 
new physics models beyond the SM. This quark level interaction 
is responsible for the purely leptonic decay $B_s \rightarrow \mu^+ 
\mu^-$ and also the semi-leptonic decays $B \rightarrow (K, K^*) 
\mu^+ \mu^-$. The semi-leptonic decays have been  
observed by BaBar and Belle \cite{babar-03, babar-06, belle-03} with 
the following branching ratios: 
\begin{eqnarray} 
B(B\rightarrow K \mu^{+} \mu^{-}) & = & 
(5.7^{+2.2}_{-1.8})\times10^{-7}, \nonumber \\ 
B(B\rightarrow K^{*}\mu^{+}\mu^{-}) & = & 
(11.0_{-2.6}^{+2.99})\times10^{-7}. 
\end{eqnarray} 
These values are close to the SM predictions 
\cite{ali-02,lunghi,kruger-01}. However there is about $20 \%$ 
uncertainty in these predictions mainly due to the errors in the 
determination of the hadronic form factors and the CKM matrix 
element $|V_{ts}|$. 
 
The decay $B_s \to \mu^+ \mu^-$ is highly suppressed in SM. Its 
branching ratio is predicted to be $(3.35\pm0.32)\times10^{-9}$ 
\cite{blanke,buchalla,buras-01}. This decay is yet to be observed  
experimentally. Recently the upper bound on its branching ratio  
has been improved to \cite{cdf-07} 
\begin{equation} 
B(B_s \rightarrow \mu^+ \mu^-) < 5.8 \times10^{-8} 
\quad  (95\% ~{\rm C.L.}) \; , 
\label{mumu-lim} 
\end{equation} 
which is still more than an order of magnitude  
above its SM prediction. 
$B_{s}\rightarrow\mu^{+}\mu^{-}$ will be one 
of the important rare $B$ decays to be studied at the upcoming Large 
Hadron Collider (LHC) and we expect that the sensitivity of the 
level of the SM prediction can be reached 
with $\sim 1$ fb$^{-1}$ of data.
\cite{schneider,maria}. 
 
Many new physics models predict an order of magnitude  
enhancement or more in $B(B_s \to \mu^+ \mu^-)$. These include theories 
with $Z^{'}$ mediated vector bosons \cite{london-97}, as well as 
multi-Higgs doublet models that violate \cite{london-97} or obey 
\cite{hewett-89} natural flavor conservation. In 
\cite{alok-sankar01}, it was shown that the new physics mediated by 
vector bosons is highly constrained by the measured values 
of the branching ratio of $B \rightarrow (K,K^*) \mu^+ \mu^-$. 
As a result, an order of magnitude enhancement 
in $B(\bsll)$ from new physics vector or axial vector operators is ruled 
out. On the other hand, such an enhancement from the  
scalar/pseudoscalar new physics (SPNP) operators is still allowed, since the 
most stringent bound on the SPNP operators comes from $B(\bsll)$ 
itself. In particular, multi-Higgs doublet models or supersymmetric 
(SUSY) models with large $\tan \beta$ can give rise to such an enhancement.

Apart from the branching ratios of the purely leptonic 
and semi-leptonic decays, there are other observables which are 
sensitive to the SPNP contribution to $b \to s$ transitions. These 
are forward-backward (FB) asymmetry $A_{FB}$
of muons \cite{ali-92} in 
$\bkll$ and longitudinal polarization (LP) asymmetry 
$A_{LP}$ of muons in $\bsll$ \cite{handoko-02}.
Both these are predicted to be zero in the SM.
Therefore, any nonzero measurement of one of these
asymmetries is a signal for new physics.
In addition, these asymmetries are almost independent of
form factors and CKM matrix element uncertainties,
which makes them attractive candidates in searches
for new physics. 
In this paper we investigate what constraints the 
recently improved upper bound on 
$B(B_s\rightarrow  \mu^{+}\mu^{-})$ 
puts on the possible SPNP contribution to 
$A_{FB}$ and $A_{LP}$.
Do SPNP operators enhance 
these observables to  sufficiently large values to be measurable in 
future experiments? 
 
The paper is organized as follows. 
In section \ref{afb}, we study the 
effect of possible SPNP contribution to $A_{FB}$.
In section \ref{lp}, we calculate the possible $A_{LP}$
enhancement due to SPNP, and point out
some interesting experimental possibilities. 
In section \ref{concl}, we present our
conclusions. 
 
%%%%%%%%%%%%%%%%%%%%%%%%%%%%%%%%%%%%%%%%%%%%%%%%%%%%%%%%%%%%% 
\section{Forward-backward asymmetry  in $\bkll$} 
\label{afb} 
 
There are numerous studies in literature of the FB asymmetry of leptons  
in the SM and its possible extensions 
\cite{ali-00,yan-00,bobeth-01,erkol-02,demir-02,li-04,chen-05}.  
In the SM, the FB asymmetry of muons in $B \rightarrow K \mu^+ \mu^-$ 
vanishes (or to be more precise, is negligibly small) because the  
hadronic current for $B \rightarrow K$ transition does not 
have any axial vector contribution. However this asymmetry can be nonzero in multi-Higgs  
doublet models and supersymmetric models with large $\tan \beta$,
due to the contributions from Higgs bosons. Therefore FB asymmetry 
in $B \rightarrow K \mu^+ \mu^-$ is expected to serve as an important probe to 
test the existence and importance of an extended Higgs sector \cite{erkol-02,chen-05}. Any 
nonzero measurement of this asymmetry will be a clear signal of new physics. 
 
The average (or integrated) FB asymmetry of muons in $B\rightarrow K 
\mu^+ \mu^-$, which is denoted by $\left\langle A_{FB}\right\rangle$, has been measured 
by BaBar \cite{babar-06} and Belle \cite{belle-06, ikado-06} to be 
\begin{equation} 
\left\langle A_{FB}\right\rangle  =  (0.15_{-0.23}^{+0.21} \pm 0.08)\, 
\, \, \, \, \, 
 {\rm (BaBar)} \, , 
\end{equation} 
\begin{equation} 
\left\langle A_{FB}\right\rangle  = (0.10 \pm 0.14 \pm 0.01) \, \, 
\, \, {\rm (Belle)}. \label{fb_exp1} 
\end{equation} 
These measurements are consistent with zero. But on the other hand,  
they can be as high as $\sim 40\%$ within $2\sigma$ error bars. 
 
\subsection{Calculation of $A_{FB}$} 

We consider new physics  in the form of scalar/pseudoscalar 
operators. 
The effective Lagrangian for the quark level 
transition $b \rightarrow s \mu^+ \mu^-$ can be written as 
\begin{equation} 
L(b \rightarrow s \mu^{+} \mu^{-}) = L_{SM} + L_{SP}\,, 
\label{lag-tot} 
\end{equation} 
where 
\begin{equation} 
\begin{split} 
L_{SM} &= \frac{\alpha G_F}{\sqrt{2} \pi} V_{tb} V^\star_{ts} 
\biggl\{ C^{\rm eff}_9        (\bar{s} \gamma_\mu P_L b)    \, 
\bar{\mu} \gamma_\mu \mu  + 
%&+ 
C_{10}              (\bar{s} \gamma_\mu P_L b)        \,   \bar{\mu} \gamma_\mu \gamma_5 \mu \\ 
&- 2 \frac{C^{\rm eff}_7}{q^2} m_b \, (\bar{s} i \sigma_{\mu\nu} 
q^\nu P_R b) \, \bar{\mu} \gamma_\mu \mu \biggr\} \, , 
\end{split} 
\label{SML} 
\end{equation} 
\begin{equation} 
L_{SP}  =  \frac{\alpha G_F}{\sqrt{2} \pi} V_{tb} V^\star_{ts} 
\biggl\{ R_S \, (\bar{s}\,P_R\, b) \, \bar{\mu} \,\mu + R_P \, 
(\bar{s}\,P_R\, b) \, \bar{\mu} \gamma_5 \mu \biggr\} \; . 
\label{LSP} 
\end{equation} 
Here $P_{L,R} = (1 \mp \gamma_5)/2$ and $q_\mu$ is the sum of $4$-momenta of 
$\mu^+$ and $\mu^-$. $R_S$ and $R_P$ are the new physics
scalar and pseudoscalar  couplings respectively. 
In our analysis we assume that there are no additional CP phases 
apart from the single CKM phase. Under this assumption, $R_S$ and 
$R_P$ are real. 
Within SM, the Wilson coefficients in eq.~(\ref{SML})  
have the following values: 
\begin{equation} 
C_{7}^{\rm eff} = -0.310 \; , \;  
C_{9}^{\rm eff} = +4.138 + Y(q^2) \; , \; C_{10} = -4.221\;, 
\end{equation} 
where the function $Y(q^2)$ is given in \cite{buras-95,misiak-95}.

The normalized FB asymmetry is defined as 
\begin{eqnarray} 
A_{FB}(z)= \frac{\int_0^{1}dcos\theta \frac{d^2\Gamma}{dz 
dcos\theta}-\int_{-1}^{0}dcos\theta \frac{d^2\Gamma}{dz dcos\theta}} 
{\int_0^{1}dcos\theta \frac{d^2\Gamma}{dz dcos\theta}+\int_{-1}^{0}dcos\theta \frac{d^2\Gamma}{dz 
d\cos\theta}}\;. 
\end{eqnarray} 
In order to calculate the FB asymmetry, we first need to 
calculate the differential decay width. 
The decay amplitude for $B(p) \to K(p') \mu^+(p_+) \mu^-(p_-)$ 
is given by 
\begin{eqnarray} 
M\,(B\rightarrow K \mu^{+}\mu^{-}) &=& \frac{\alpha G_F}{2\sqrt{2} \pi} V_{tb} V^\star_{ts}   
\nonumber \\ 
&\times& 
\Bigg[\left< K(p') \left|\bar{s}\gamma_{\mu}b\right|B(p)\right> 
\left\{C_{9}^{\rm eff}\bar{u}(p_+)\gamma_{\mu}v(p_-)  
+C_{10}\bar{u}(p_+)\gamma_{\mu}\gamma_{5} v(p_-)\right\}  
\nonumber \\ 
& & - 2\frac{C^{\rm eff}_7}{q^2} m_b \left< K(p')\left|\bar{s}
i\sigma_{\mu\nu}q^{\nu}b\right|B(p)\right> 
\bar{u}(p_+)\gamma_{\mu}v(p_-) 
\nonumber \\ 
& & + \left< K(p')\left|\bar{s}b\right|B(p)\right> 
\left\{R_S\bar{u}(p_+)v(p_-)+ 
R_P\bar{u}(p_+)\gamma_{5}v(p_-)\right\} \Bigg]\; ,
\end{eqnarray} 
where $q_\mu = (p-p')_\mu = (p_+ + p_-)_\mu$.
The relevant matrix elements are 
\begin{eqnarray} 
\left< K(p') \left|\bar{s}\gamma_{\mu}b\right| 
B(p)\right>&=&(2p-q)_{\mu}f_{+}(z)+(\frac{1-k^2}{z})\, 
q_{\mu}[f_{0}(z)-f_{+}(z)]\;, 
\end{eqnarray} 
\begin{eqnarray} 
\left< K(p')\left|\bar{s}i\sigma_{\mu\nu}q^{\nu}b\right| 
B(p)\right>=-\Big[(2p-q)_{\mu}q^2-(m_{B}^{2}-m_{K}^{2})q_{\mu}\Big]\,\frac{f_{T}(z)}{m_B+m_{K}}\;, 
\end{eqnarray} 
\begin{eqnarray} 
\left< K(p')\left|\bar{s}b\right| 
B(p)\right>=\,\frac{m_B(1-k^2)}{\hat{m}_b}\,f_{0}(z)\; .
\end{eqnarray} 
Here, $k \equiv m_K/m_B$, $z \equiv q^2/m_{B}^{2}$ and 
$\hat{m}_b \equiv m_b/m_B$. 
In this paper, we approximate $\hat{m}_b$ by $1$. 
 
Using the above matrix elements, the double differential decay 
width can be calculated as 
\begin{eqnarray} 
\frac{d^{2}\Gamma}{dzdcos\theta} & = & \,\frac{G_{F}^{2}\alpha^{2}}{2^{9}\pi^{5}}\,|V_{tb}V_{ts}^{*}|^{2}\, m_{B}^{5}\,\phi^{1/2}(1,k^2,z)\,\beta_{\mu}\nonumber \\ 
 & \times & \Bigg[\Big(\left|A\right|^{2}\beta_{\mu}^{2}+\left|B\right|^{2}\Big)z+\frac{1}{4}\phi(1,k^2,z)\Big(\left|C\right|^{2}+\left|D\right|^{2}\Big)(1-\beta_{\mu}^{2} \cos^{2}\theta)\nonumber \\ 
 &  & + 2\hat{m_{\mu}}(1-k^{2}+z)Re(BC^{*}) +
4\hat{m_{\mu}}^{2}\left|C\right|^{2}\nonumber \\ 
 &  &
+2\hat{m_{\mu}}\,\phi^{\frac{1}{2}}(1,k^2,z)\,\beta_{\mu}\, Re(AD^{*})\, \cos\theta \Bigg] \;, 
\label{double_drate} 
\end{eqnarray} 
where 
\begin{eqnarray} 
A & \equiv & \frac{\,1}{2}(1-k^{2})f_{0}(z)R_S\;, \nonumber \\ 
B &\equiv &-\hat{m_{\mu}}C_{10}\left\{f_{+}(z)-\frac{1-k^{2}}{z}(f_{0}(z)-f_{+}(z))\right\}+ 
\frac{1}{2}(1-k^{2})f_{0}(z)R_P\;, \nonumber \\ 
C &\equiv & C_{10}\,f_{+}(z)\;, \nonumber \\ 
D &\equiv & C_{9}^{eff}\, f_{+}(z)+2\, C_{7}^{eff}\,\frac{f_{T}(z)}{1+k}\;, \nonumber \\ 
\phi(1,k^2,z) & \equiv& 1+k^{4}+z^{2}-2(k^{2}+k^{2}z+z)\;, 
\nonumber \\ 
\beta_{\mu} & \equiv & (1-\frac{4\hat{m_{\mu}}^{2}}{z})\;. 
\end{eqnarray} 
Also, $\hat{m}_\mu=m_{\mu}/m_B$ and $\theta$ is the angle between the 
momenta of $K$ meson and $\mu^-$ in the dilepton 
centre of mass frame. The kinematical variables are bounded as 
\begin{eqnarray} 
-1\leq \cos\theta\leq1\;,\nonumber\\ 
4\hat{m}^2_{\mu}\leq z \leq(1-k)^2\; .\nonumber 
\end{eqnarray} 

The form factors $f_{+,0,T}$ can be calculated in the
light cone QCD approach. Their $q^2$ dependence 
is given by \cite{ali-00} 
\begin{eqnarray} 
f(z)=f(0)\,\exp(c_1z+c_2z^2+c_3z^3)\;, 
\end{eqnarray} 
where the parameters $f(0), c_1$, $c_2$ and $c_3$ for each form 
factor are given in Table~\ref{ff-table}.  
The FB asymmetry arises from the $\cos\theta$ term in 
the last line of eq.~(\ref{double_drate}).  

\begin{table}[h] 
$$ 
\begin{array}{|l c c c c|} 
\hline 
    & \phantom{-}f(0)  &\phantom{-}c_1  & \phantom{-}c_2  & \phantom{-}c_3\\ \hline 
	f_{+}&\phantom{-}0.319^{+0.052}_{-0.041}  &\phantom{-}1.465  &\phantom{-}0.372 
&\phantom{-}0.782\\ \hline f_0 &\phantom{-}0.319^{+0.052}_{-0.041}   &\phantom{-}0.633 
&\phantom{-}-0.095  &\phantom{-}0.591\\ \hline f_T &\phantom{-}0.355^{+0.016}_{-0.055} 
&\phantom{-}1.478 & \phantom{-}0.373   &\phantom{-}0.700 \\ \hline 
\end{array} 
$$ 
\caption{Form factors for the $B \to K$ transition \cite{ali-00}.
\label{ff-table}} 
\end{table}

The calculation of FB asymmetry gives 
\begin{equation} 
A_{FB}(z)=\frac{2 \Gamma_{0} \, \hat{m_{\mu}} \, a_{1}(z) \, 
\phi(1,k^2,z) \, \beta_{\mu}^{2} \, R_S}{d\Gamma/dz}\;, \label{afb1} 
\end{equation} 
where 
\begin{equation} 
\Gamma_{0}=\frac{G_{F}^{2}\alpha^{2}}{2^{9}\pi^{5}}\,|V_{tb}V_{ts}^{*}|^{2}\, 
m_{B}^{5} \;, 
\end{equation} 
\begin{equation} 
a_{1}(z)=\frac{1}{2}(1-k^{2})C_{9}f_{0}(z)f_{+}(z)+(1-k)C_{7}f_{0}(z)f_{T}(z)\;, 
\end{equation} 
\begin{eqnarray} 
\frac{1}{\Gamma_{0}} \frac{d\Gamma}{dz} &=& \frac{1}{2}(1-k^{2})\, 
\beta_{\mu} \, \phi^{\frac{1}{2}} \, z \, f_{0}^{2}(z) \, (R_{P}^{2} + 
\beta_{\mu}^{2} R_{S}^{2}) 
\nonumber\\ 
&+& 
2(1-k^{2})\,\hat{m_{\mu}}\,C_{10}\,f_{0}(z)f_{+}(z)\,\beta_{\mu}\,\phi^{\frac{1}{2}}(z)\,(1-k^{2}+z)R_P 
\nonumber\\ 
&-& 
2(1-k^{2})\,\hat{m_{\mu}}\,C_{10}\,\beta_{\mu}\,z\,\phi^{\frac{1}{2}}f_{0}(z)\left\{ 
f_{+}(z)-\frac{1-k^{2}}{z}(f_{0}(z)-f_{+}(z))\right\} R_P 
\nonumber\\ 
&+&2\hat{m}_{\mu}^{2}\,C_{10}^{2}\,\beta_{\mu}\,\phi^{\frac{1}{2}}(z)\left\{ f_{+}(z)-\frac{1-k^{2}}{z}(f_{0}(z)-f_{+}(z))\right\}^{2} \nonumber\\ 
&+& 8\hat{m}_{\mu}^{2} \, C_{10}^{2} \, \beta_{\mu} \, 
\phi^{\frac{1}{2}}(z)f_{+}^{2}(z) 
\nonumber\\ 
&+& 
\frac{1}{3}(1+\frac{2\hat{m}_{\mu}^{2}}{z})\beta_{\mu}\phi^{\frac{3}{2}}(z) \times  
\nonumber\\ 
&& \left\{(C_{10}^{2}+C_{9}^{eff2})f_{+}^{2}(z) + \frac{4C_{7}^{eff2}}{(1+k)^{2}}f_{T}^{2}(z)+ 
\frac{4C_{9}^{eff}C_{7}^{eff}}{(1+k)}f_{+}(z)f_{T}(z)\right\} 
\nonumber\\ 
&-& 
4\hat{m_{\mu}}^{2}C_{10}^{2}f_{+}(z)\,\beta_{\mu}\,(1-k^{2}+z)\,\phi^{\frac{1}{2}}(z) \times 
\nonumber\\ 
&& \left\{f_{+}(z)-\frac{1-k^{2}}{z}(f_{0}(z)-f_{+}(z))\right\}\;. 
\end{eqnarray} 
 
From eq.~(\ref{afb1}), it is clear that $A_{FB}(z)$ is proportional 
to $\hat{m}_{\mu} (\approx 0.02)$, 
and to the scalar new physics coupling $R_S$.  
In the minimal supersymmetric standard model (MSSM) and 
two Higgs doublet models, $R_S$ itself 
is proportional to $\hat{m}_{\mu}$ and $\tan^2\beta$. 
Hence a large FB asymmetry is possible only 
for exceptionally large values of $\tan\beta$. 
 
The average FB asymmetry is obtained by 
integrating the numerator and denominator of eq.~(\ref{afb1}) separately over dilepton 
invariant mass, which leads to 
\begin{equation} 
\langle A_{FB} \rangle =
\frac{2 \Gamma_{0} \, \hat{m}_{\mu} \, 
\beta_{\mu}^{2} \, R_S \,
\int dz \, a_{1}(z) \, \phi(1,k^2,z)}{\Gamma(B \to K \mu^+ \mu^-)} =
\frac{2 \tau_B \Gamma_{0} \, \hat{m}_{\mu} \, 
\beta_{\mu}^{2} \, R_S \,
\int dz \, a_{1}(z) \, \phi(1,k^2,z)}{B(B \to K \mu^+ \mu^-)} \; .
\label{avg-afb}
\end{equation}
where $B(B \to K \mu^{+} \mu^{-})$ is the total branching ratio of 
$B \to K \mu^{+} \mu^{-}$. 
The numerator in eq.~(\ref{avg-afb}) can be calculated to be
\begin{equation}
2 \tau_B \Gamma_{0} \, \hat{m_{\mu}} \, 
\beta_{\mu}^{2} \, R_S \,
\int dz \, a_{1}(z) \, \phi(1,k^2,z) =
(5.25 \times 10^{-9})(1 \pm 0.20)R_S \; ,
\end{equation}
whereas the total branching ratio,
including the contribution of SPNP operators, 
is given by \cite{bobeth-01} 
\begin{equation} 
B(B \rightarrow K \mu^+ \mu^-)  =   
\left[5.25 + 0.18(R_S^2 + R_P^2)  
-0.13 R_P \right] \, (1 \pm 0.20) \times 10^{-7}\;. 
\label{brsl} 
\end{equation} 
In the SM calculation of $B(\bkll)$, two vector form factors,
$f_0$ and $f_+$, as well as the tensor form factor $f_T$
appear. The SPNP contribution, on the other hand,
is only through $f_0$.
We have made the assumption that the fractional uncertainties 
in all the form factors are the same. 
The $|V_{ts}|$ dependence in the numerator and denominator  
of eq.~(\ref{avg-afb}) cancels completely, whereas 
the errors due to the form factors uncertainties cancel partially. 
We conservatively take the net error in 
$\left\langle A_{FB}\right\rangle$ to be $30\%$, leading to 
\begin{equation} 
\left\langle A_{FB}\right\rangle =\frac{5.25 \times10^{-9}\,
R_S}{\left[5.25 + 0.18(R_S^2 + R_P^2)  
-0.13 R_P\right]\times 10^{-7}} \, (1 \pm 0.3) \;. \label{fb_np1} 
\end{equation} 
\begin{table} 
\begin{center} 
\begin{displaymath} 
\begin{tabular}{|l|l|} 
\hline \spp $G_F = 1.166 \times 10^{-5} \; \gev^{-2}$ & 
     $m_{B_s}=5.366 \; \gev$ \\ 
\spp $\alpha = 7.297 \times 10^{-3}$ & 
     $ m_B=5.279 \; \gev$  \\ 
\spp $\tau_{B_s} = (1.437_{-0.030}^{+0.031}) \times 10^{-12} s$  & 
     $V_{tb}= 1.0 $  \\ 
\spp $\tau_{B_d} = 1.53 \times 10^{-12} s$  & 
     $V_{ts}= (40.6 \pm 2.7) \times 10^{-3}$ \\ 
\spp $m_{\mu}=0.105 \;\gev$ & 
     $f_{B_s}=(0.259 \pm 0.027) \; \gev$ \cite{mackenzie}\\ 
\spp $m_K= 0.497 \; \gev$ & 
     \\ \hline 
\end{tabular} 
\end{displaymath} 
\caption{Numerical inputs used in our 
  analysis. Unless explicitly specified, they are taken from the 
  Review of Particle Physics~\cite{Yao:2006px}.\label{tab:inputs}} 
\end{center} 
\end{table} 

\subsection{Constraints on $\langle A_{FB} \rangle$ from $B(\bsll)$}

We now want to see what constraints the present upper bound 
on $B(\bsll)$ puts on the maximum possible value of 
$\left\langle A_{FB}\right\rangle$. The present experimental upper 
limit on $B(B_s \to \mu^+ \mu^-)$ is an order of magnitude larger 
than the SM prediction. In such a situation, the SM amplitude for 
this decay will be much smaller than the new physics amplitude and 
hence can be neglected in determining the constraints on new physics 
couplings, $R_S$ and $R_P$. 
In other words, we will assume that SPNP  
operators  saturate the present upper limit.  
Therefore we need to consider only the contribution of 
$L_{SP}$ to the decay rate of  $B_{s}\rightarrow\mu^{+}\mu^{-}$.  
 
The decay amplitude for $\bsll$ is given by 
\begin{equation} 
M\,(B_{s}\rightarrow \mu^{+}\mu^{-})\,=\,\frac{\alpha G_F}{2\sqrt{2} \pi} V_{tb} V^\star_{ts} 
\langle 0\left|\overline{s}\gamma_{5}b\right|B_{s}\rangle 
\left[R_S\bar{u}(p_{\mu})v(p_{\bar{\mu}})+ 
R_P\bar{u}(p_{\mu})\gamma_{5}v(p_{\bar{\mu}})\right]\;. 
\end{equation} 
On substituting 
\begin{equation} 
\langle0\left|\overline{s}\gamma_{5}b\right|B_{s}\rangle\,=\,-i\frac{f_{B_{s}}m_{B_{s}}^{2}}{m_{b}+m_{s}}\;, 
\end{equation} 
we get 
\begin{equation} 
M\,(B_{s}\rightarrow \mu^{+}\mu^{-})\,=\,-i\,\frac{\alpha G_F}{2\sqrt{2} \pi} V_{tb} V^\star_{ts} 
\frac{f_{B_{s}}m_{B_{s}}^{2}}{m_{b}+m_{s}} 
\left[R_S\bar{u}(p_{\mu})v(p_{\bar{\mu}})+ 
R_P\bar{u}(p_{\mu})\gamma_{5}v(p_{\bar{\mu}})\right]\;, 
\end{equation} 
where $m_{b}$ and $m_{s}$ are the masses of bottom and strange quark, respectively. 
The calculation of the branching ratio $B(\bsll)$ gives 
\begin{equation} 
B(B_{s}\rightarrow \mu^{+}\mu^{-})\,=\, \frac{G_{F}^{2}\alpha^{2} 
m_{B_s}^{3} \tau_{B_s}}{64\pi^{3}} |V_{tb}V_{ts}^{*}|^{2} \, 
f_{B_{s}}^{2} \, (R_{S}^{2} + R_{P}^{2})\;. 
\end{equation} 
Here we have neglected terms of order $m_s/m_b$ and approximated 
$m_{B_s}/m_{b}$ by $1$. Taking $f_{B_s}=(0.259 \pm 0.027)\, \gev$, we 
get 
\begin{equation} 
B(B_{s}\rightarrow \mu^{+}\mu^{-})\,=\,(1.43 \pm 0.30) \times 
10^{-7}\, (R_{S}^{2} + R_{P}^{2}) \,. 
\label{br_lep} 
\end{equation} 
Equating the expression in eq.~(\ref{br_lep}) to the present 
95\% C.L. upper limit in eq.~(\ref{mumu-lim}),  
we get the inequality 
\begin{equation} 
(R_{S}^{2} + R_{P}^{2}) \leq 0.70 \; , 
\label{lepconst} 
\end{equation} 
where we have taken the $2\sigma$ lower bound for the  
coefficient in eq.~(\ref{br_lep}). 
Thus, the allowed region in the $R_S$--$R_P$ parameter 
space is the interior of a circle of radius $ \approx 0.84$ 
centered at the origin. 
 
In \cite{alok_amol_uma}, it was shown that  
the SPNP operators cannot lower  
$B(B \to K \mu^+ \mu^-)$ below its SM prediction. 
Therefore from eq.~(\ref{fb_np1}), 
the maximum value of $\left\langle A_{FB}\right\rangle$ 
with the current upper bound on $B(\bkll)$ is $1.34\%$ at $2\sigma$. 
If $B(\bsll)$ is bounded to $10^{-8}$, 
the $2\sigma$ maximum value of  
$\left\langle A_{FB}\right\rangle$ will be $0.56\%$.

A naive estimation suggests that the measurement of an asymmetry $\left\langle A_{FB}\right\rangle$ of a 
decay with the branching ratio $\cal B$ at $n \sigma$ C.L. with only 
statistical errors require 
\begin{equation} 
N \sim \left( \frac {n}{{\cal B} \,\left\langle A_{FB}\right\rangle}
\right)^2 \; 
\end{equation} 
number of events. 
For $B \rightarrow K \mu^+ \mu^-$, if $\left\langle 
A_{FB}\right\rangle$ is $1\%$ at $2\sigma$ C.L., then the required number of events 
will be as high as $10^{18}$ ! Therefore it is very difficult to observe 
such a low value of FB asymmetry in experiments. Hence FB asymmetry 
of muons in $B \rightarrow K \mu^+ \mu^-$ will play no role in testing SPNP.

%%%%%%%%%%%%%%%%%%%%%%%%%%%%%%%%%%%%%%%%%%%%%%%%%%%%%%%%%%%%% 
\section{Longitudinal polarization asymmetry in $\bsll$ } 
 
\label{lp} The longitudinal polarization asymmetry of muons in $B_s 
\to \mu^+ \mu^-$ is a clean observable that depends only on SPNP 
operators. It vanishes in the SM, whereas its value is nonzero if 
and only if the new physics contribution is in the form of scalar 
operator. Therefore any nonzero measurement of this observable 
$A_{LP}$ will confirm the existence of an 
extended Higgs sector. The observable $A_{LP}$ was introduced in 
ref. \cite{handoko-02}, though the corresponding analysis 
in the context 
of $K_L \to \mu^+ \mu^-$ had been carried out earlier 
\cite{botella,kll,geng,ecker}.  
In this section, we will determine the allowed values of $A_{LP}$ 
consistent with the present upper bound on $B(B_s \to \mu^+ \mu^-)$, 
and explore the correlation between these two quantities.

The most general model independent form of the effective 
Lagrangian for the quark level transition $b \to s \mu^+ \mu^-$ that 
contributes to the decay $B_s \to \mu^+ \mu^-$ has the form \cite{fkmy,guetta} 
\begin{eqnarray} 
  L & = & \frac{G_F \alpha}{2 \sqrt{2} \pi} 
  \left( V_{ts}^\ast V_{tb} \right) \, 
  \left\{ 
  R_A 
  (\bar{s} \, \gamma_\mu \gamma_5 \, b) 
  (\bar{\mu} \, \gamma^\mu \gamma_5 \, \mu) 
  \right. \nonumber \\ 
  & & \left. \; \; \; \; \; \; \; \; \; \; \; \; \; \; 
  + 
  \fps 
  (\bar{s} \, \gamma_5 \, b) 
  (\bar{\mu} \, \mu) 
  + 
  \fpp 
  (\bar{s} \, \gamma_5 \, b) 
  (\bar{\mu} \, \gamma_5 \, \mu) 
  \right\} \; , 
  \label{eqn:heff1} 
\end{eqnarray} 
where $R_P, R_S$ and $R_A$ are the strengths of the scalar, 
pseudoscalar and axial vector operators respectively.  
Note that the effective Lagrangian in eq.~(\ref{eqn:heff1}) is essentially  
the same as the effective Lagrangian given in eq.~(\ref{lag-tot}). Here we have dropped 
$C_7$ and $C_9$ terms which do not contribute to $\bsll$. In addition, 
the $R_A$ in eq.~(\ref{eqn:heff1}) is the sum of SM and new physics contributions. 
 
In SM, the 
scalar and pseudoscalar couplings $\fps^{\rm SM}$ and $\fpp^{\rm 
SM}$ receive contributions from the penguin diagrams with physical 
and unphysical neutral scalar exchange and are highly suppressed: 
\begin{equation} 
 \fps^{\rm SM} = \fpp^{\rm SM} \propto \frac {(m_{\mu} m_b)}{m_{W}^{2}} \sim 10^{-5} \; . 
\end{equation} 
Also, $\faa^{\rm SM } = {Y(x)}/{\sin^2 \theta_W}$, where $Y(x)$ is 
the Inami-Lim function \cite{inamilim} 
\begin{equation} 
Y(x)=\frac{x}{8}\, \left[\frac{x-8}{x-1}+ \frac{3x}{(x-1)^2} \ln x 
\right] \; , 
\end{equation} 
with $x =({m_t}/{M_W})^2$. Thus, $\faa^{\rm SM }\simeq4.3$.

The calculation of the branching ratio gives \cite{handoko-02,fkmy} 
\begin{equation} 
B(\bqll) = a_s \left[ 
  \left| 2 m_\mu \faa  - \frac{m_{B_s}^2}{m_b + m_s} \fpp\right|^2 
  + \left( 1 - \frac{4 m_{\mu}^{2}}{m_{B_s}^2}\right) 
  \left| \frac{m_{B_s}^2}{m_b + m_s} \fps\right|^2 
   \right]\;, 
  \label{blep_gen} 
\end{equation}  
where 
\begin{equation} 
a_s \equiv \frac{G_F^2 \alpha^2}{64 \pi^3} \, 
  \left| V_{ts}^\ast V_{tb}  \, \right|^2 \tau_{B_s} f_{B_s}^2 m_{B_s} \, 
 \sqrt{ 1 - \frac{4 m_\mu^2 }{m_{B_s}^2} } \; . 
\end{equation} 
Here $\tau_{B_s}$ is the lifetime of $B_s$. 
Eq.~(\ref{blep_gen}) represents the most general expression for the 
branching ratio of $B_s \to \mu^+ \mu^-$. 
 
We now derive an expression for the lepton polarization. In the  
rest frame of $\mu^+$,  
we can define only one direction $\overrightarrow{p}_{-}$, the three momentum of $\mu^-$. 
The unit longitudinal polarization 4-vectors along that direction are 
\begin{eqnarray} 
\bar{s}^\mu_{\mu^{\pm}}=(0,\ \hat{e}_{L}^{\pm})=\Big(0,\ \pm 
\frac{\overrightarrow{p}_{-}}{|\overrightarrow{p}_{-}|}\Big)\; . 
 \label{smu} 
\end{eqnarray} 
Transformation of unit vectors from the rest frame of $\mu^+$ to the 
center of mass frame of leptons (which is also the rest frame of $B_s$ meson) can be accomplished by the Lorentz 
boost. After the boost, we get 
\begin{eqnarray} 
s_{\mu^{\pm}}^{\mu}=\Big(\frac{|\overrightarrow{p}_{-}|}{m_{\mu}},\ 
\pm\frac{E_{\mu}\overrightarrow{p}_{-}}{m_{\mu}|\overrightarrow{p}_{-}|}\Big)\;, 
\label{smuboost} 
\end{eqnarray} 
where $E_{\mu}$ is the muon energy. 
 
The longitudinal polarization asymmetry of muons in $B_s \to \mu^+ 
\mu^-$ is defined as 
\begin{equation} 
A^{\pm}_{LP}\ =\ \frac{\Gamma( \hat{e}_{L}^{\pm})\ -\ 
\Gamma( -\hat{e}_{L}^{\pm})}{\Gamma( 
\hat{e}_{L}^{\pm})\ +\ 
\Gamma(-\hat{e}_{L}^{\pm})} \; . 
 \label{pl} 
\end{equation} 
Thus we get \cite{handoko-02} 
\begin{eqnarray}  
\alp &=& \frac{2\sqrt{ 1- \frac{4 m_\mu^2 }{m_{B_s}^2} }   
\left[  \frac{m_{B_s}^2}{m_b + m_s} \fps \left( 2 m_\mu \faa   
- \frac{m_{B_s}^2}{m_b + m_s}~\fpp \right) \right]  } 
{\left| 2 m_\mu \faa  - \frac{m_{B_s}^2}{m_b + m_s}  \fpp \right|^2 
 + (1-\frac{4m_\mu^2}{m_{B_s}^2}) \left|  \frac{m_{B_s}^2}{m_b + m_s} \fps \right|^2 }\;, 
\label{eqn:alp} 
\end{eqnarray} 
with  $\alp^+ = \alp^- \equiv \alp$. It is clear from eq.~(\ref{eqn:alp})  
that $\alp$ can be nonzero if and only if 
$\fps \neq 0$, i.e. for $\alp$ to be nonzero, we must have 
contribution from SPNP operators. Within the SM, 
$\fps \simeq 0$ and hence $\alp \simeq 0$.  
 
Using eq.~(\ref{blep_gen}), we can eliminate $R_A$ and $R_P$ from eq.~(\ref{eqn:alp}) in 
favour of the physical observables $B(\bsll)$ and $a_s$. We get \cite{handoko-02} 
\begin{eqnarray} 
  \alp &  = & \pm \frac{ 2 a_s}{B(\bqll)} \, 
 \sqrt{ 1 - \frac{4 m_\mu^2 }{m_{B_s}^2} } 
\times
 \nonumber \\ 
  & & \frac{m_{B_s}^2\,R_S}{m_b + m_s}  
  \sqrt{\frac{B(\bqll)}{a_s} - \left( 1 - \frac{4 m_{\mu}^2 }{m_{B_s}^2} \right) 
    \left| \frac{m_{B_s}^2\,R_S}{m_b + m_s}  \right|^2 } 
   \; . 
\label{eqn:apl_br}
\end{eqnarray} 
Eq. (\ref{eqn:apl_br}) represents a general relation between the 
longitudinal polarization asymmetry $A_{LP}$
and the branching ratio of $B_s \to \mu^+ \mu^-$. 
 
We now explore the correlation between $\alp$ and $B(\bsll)$. 
It is quite obvious that when $B(\bsll) \gtrsim 10^{-8}$,  we can neglect the SM 
contribution in obtaining the bounds on $R_S$ and $R_P$. However if $B(\bsll)$ is of the 
order of the SM prediction, then we will have to take into account the SM contribution as well. 
Therefore it is reasonable to consider both the cases separately. 

\begin{figure} 
\centering 
\includegraphics[width=0.8\textwidth]{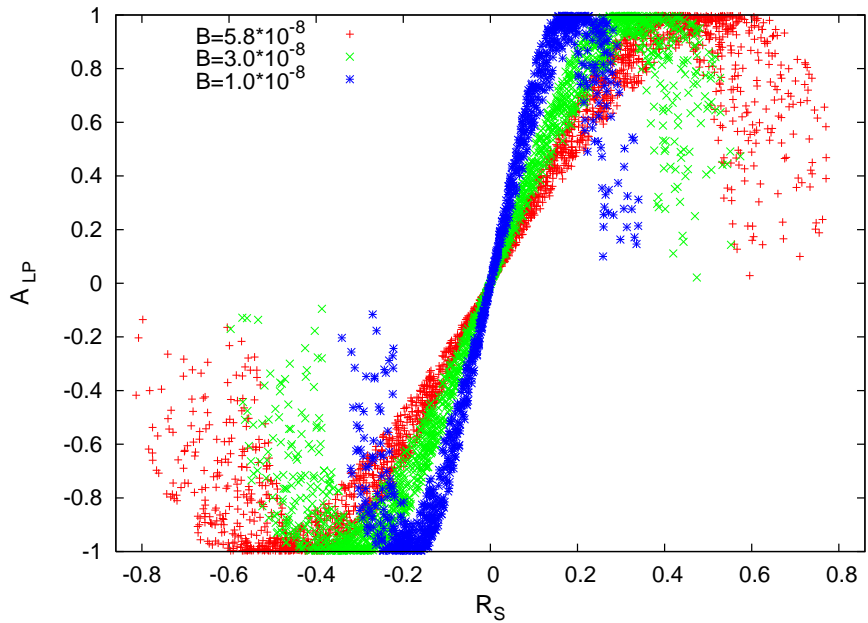} 
\caption{$A_{LP}$ vs $R_s$ plot for $B(B_s\rightarrow 
\mu^{+}\mu^{-})=(5.8,3.0,1.0) \times 10^{-8}$} 
\centering \label{fig1} 
\end{figure} 

\begin{figure} 
\centering 
\includegraphics[width=0.8\textwidth]{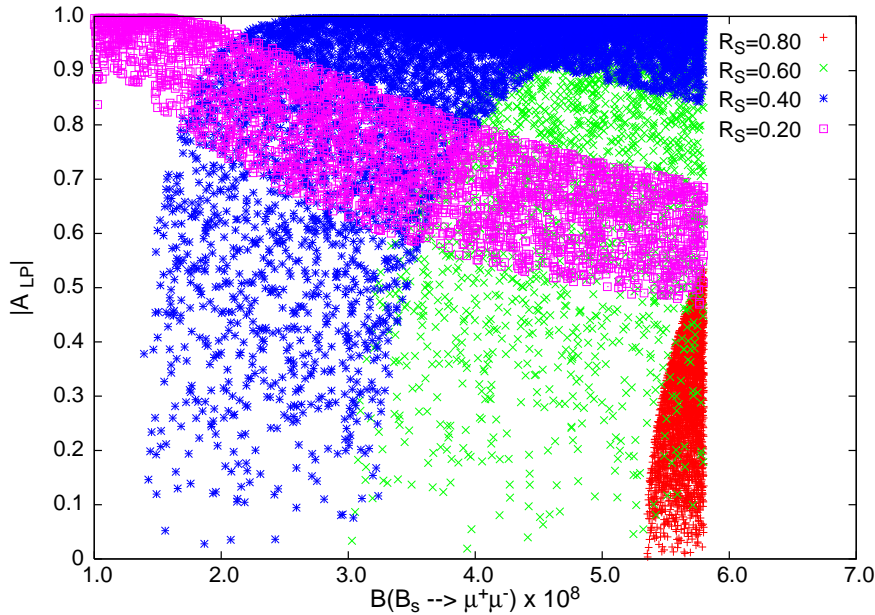} 
\caption{Plot between $|\alp|$ and $B(B_s \to \mu^+ \mu^-)$ 
for different $R_S$ values, when $B(\bsll) \gtrsim 10^{-8}$.
The region $B(\bsll) > 5.8 \times  10^{-8}$ is ruled out
by experiments to $95\%$ C.L..
} \centering   
\label{fig2} 
\end{figure} 

\subsection{$B(\bsll)  \gtrsim 10^{-8}$} 

We first consider the constraints on $\alp$ coming from the 
present upper bound on $B(B_s \to \mu^+ \mu^-)$. 
Fig.~\ref{fig1} shows the plot between  $\alp$ and $R_S$ for  
three different values of $B(\bsll) \gtrsim 10^{-8}$.  
Fig.~\ref{fig2} is a plot between 
$|\alp|$ and $B(B_s \to \mu^+ \mu^-)$ for various allowed values of 
$R_S$. The bands in Figs.~\ref{fig1} and \ref{fig2} are mainly due  
to the uncertainties in CKM matrix element $|V_{ts}|$ and decay 
constant $f_{B_s}$.

We see from Fig.~\ref{fig1} that the maximum possible value of $\alp$ 
consistent with the present upper bound on  $B(B_s \to \mu^+ \mu^-)$ 
is $100 \%$, i.e. the present upper bound of $B(B_s \to \mu^+ \mu^-)$ 
does not put any constraint on $\alp$. Indeed, $B(B_s \to \mu^+ \mu^-)$ 
will be unable to put any constraint on $\alp$ even if it is 
as low as $10^{-8}$.

Thus we see that the recently improved upper bound on the branching 
ratio of $B_s \to \mu^+ \mu^-$, which provides the most stringent 
bound on SPNP couplings, fails to put any 
bound on $\alp$. 
{\it Therefore $\alp$ is more sensitive to SPNP operators as compared 
to $B(B_s \to \mu^+ \mu^-)$.} 
Any nonzero measurement of  $\alp$ will be evidence 
for an extended Higgs sector. 
 
We would like to emphasize another important point: The measurement 
of $B(B_s \to \mu^+ \mu^-)$ will only give the 
allowed range for the values of the SPNP couplings $R_S$ and $R_P$.  
However the simultaneous determination of $B(B_s \to \mu^+ \mu^-)$ and $\alp$ will 
allow the determination of new physics scalar coupling $R_S$ 
(see Fig.~\ref{fig2}) and this in turn will enable us to determine the new 
physics pseudoscalar coupling $R_P$.

\subsection{$B(\bsll)  \lesssim 10^{-8}$} 

LHC is expected to reach the SM sensitivity in $\bsll$. 
In fact, it may
even go $5\sigma$ below the SM prediction \cite{schneider}. 
Therefore it is worth considering the case 
when $B(\bsll)$ is of the order of the SM prediction. In this section we  
study the correlation between $\alp$ and $B(B_s \to \mu^+ \mu^-)$ under 
the assumption that $B(\bsll)$ is close to its SM prediction.  
 
Taking $R_A=R_A^{SM}$, eq.~(\ref{blep_gen}) gives
\begin{equation} 
B(\bsll)=a_s\Big[28.8(R_S^2\,+\,R_P^2)\,-\,9.7R_P\,+\,0.81 \Big]\;, 
\end{equation} 
which leads to 
\begin{equation} 
R_S^2\,+\,(R_P\,-\,0.165)^2\,=\,\frac{0.035 \, B(\bsll)}{a_s}\;. 
\end{equation} 
This corresponds to a circle in $R_S-R_P$ plane with centre at $(R_S=0,\;R_P=0.165)$ and radius given 
by 
%\begin{equation}  
$r=\sqrt{0.035\,B(\bsll)/a_s}$ . 
%\end{equation} 

\begin{figure} 
\centering 
\includegraphics[width=0.8\textwidth]{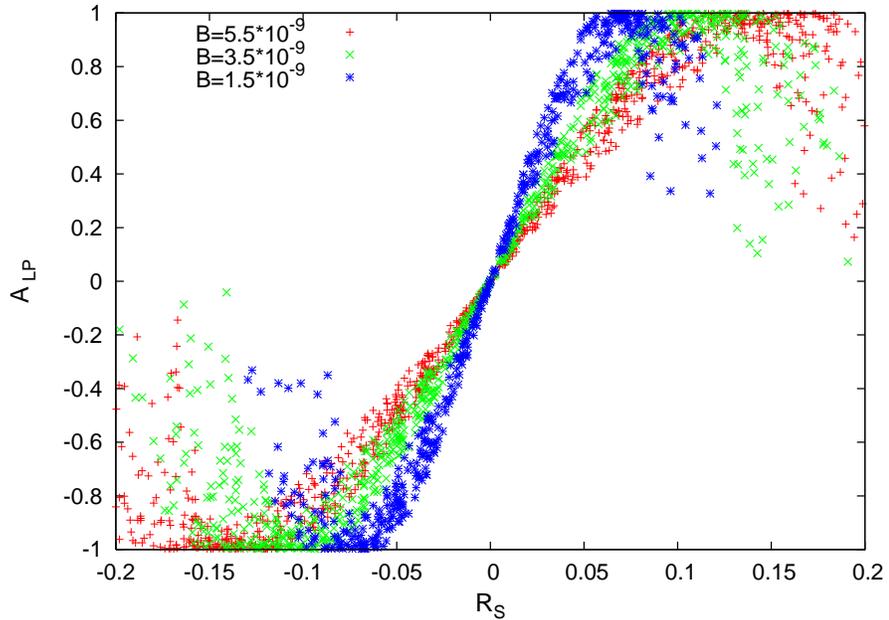} 
\caption{$A_{LP}$ vs $R_s$ plot for $B(B_s\rightarrow 
\mu^{+}\mu^{-})=(5.5,3.5,1.5) \times 10^{-9}$} 
 \centering  
\label{fig3} 
\end{figure} 

\begin{figure} 
\centering 
\includegraphics[width=0.8\textwidth]{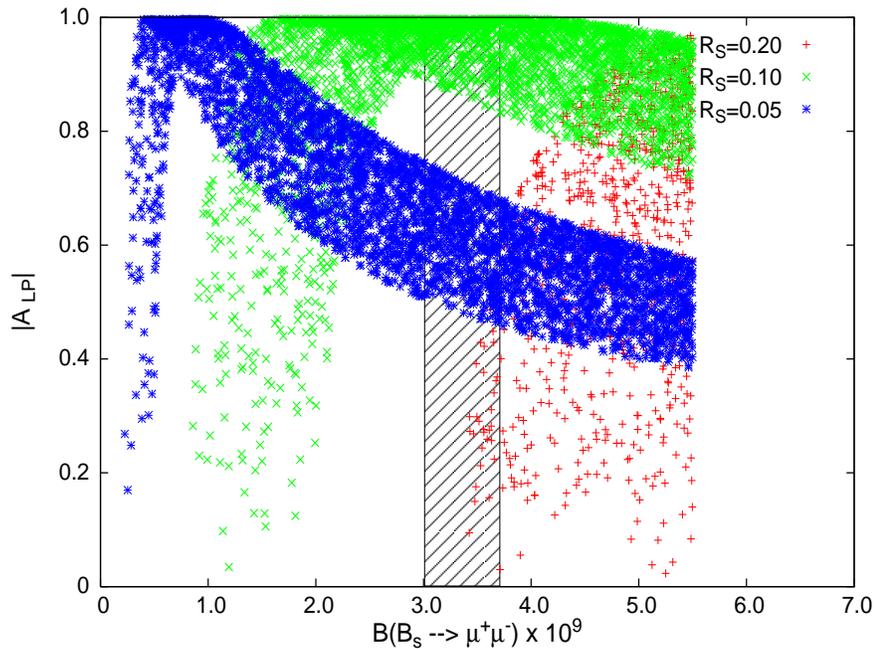} 
\caption{Plot between $|\alp|$ and $B(B_s \to \mu^+ \mu^-)$ 
for different $R_S$ values, when $B(\bsll) \lesssim 10^{-8}$.
The vertical shaded band corresponds to $1\sigma$ theoretical
prediction within the SM. 
The region $B(\bsll) \gtrsim 5 \times 10^{-9}$ is not ruled out; 
here we just concentrate on the region near the SM prediction.
} \centering  
\label{fig4} 
\end{figure} 
 
Fig.~\ref{fig3} shows the plot between  $\alp$ and $R_S$ for  
three different values of $B(\bsll)\lesssim  10^{-8}$.  
Fig.~\ref{fig4} is a plot between 
$|\alp|$ and $B(B_s \to \mu^+ \mu^-)$ for various allowed values of 
$R_S$. 
It is obvious from fig.~\ref{fig3}  
that $\alp$ can be $100\%$ even if $B(B_s \to \mu^+ \mu^-)$ 
is close to its SM prediction. 
 
We now consider three exciting experimental possibilities,
all of which can be accounted for with SPNP.

\subsubsection{$B(\bsll)$ is consistent with SM but $\alp\neq 0$}  

It is possible to have a non-zero value of $\alp$ even if 
$B(B_s \to \mu^+ \mu^-)$ is equal to its SM prediction. 
We can re-write eq.~(\ref{blep_gen}) in the following form: 
\begin{equation} 
B(\bsll)=a_s[(b_{SM}-b_P)^2\,+\,b_{S}^2]\;, 
\end{equation} 
where 
\begin{equation} 
b_{SM}=2 m_\mu R_A^{SM}\;, \quad
%\end{equation} 
%\begin{equation} 
b_{P}=\frac{m_{B_s}^2 } {m_b + m_s} R_P\;, \quad
%\end{equation} 
%\begin{equation} 
b_{S}=\sqrt{1 - \frac{4 m_\mu^2 }{m_{B_s}^2}}  
\frac{m_{B_s}^2}{m_b + m_s} R_S\;. 
\label{b-def}
\end{equation} 
Here we have taken $R_A=R_A^{SM}$, i.e. we have considered 
new physics only through the SPNP operators.
Now if $B(B_s \to \mu^+ \mu^-)$ is equal to its SM prediction, then 
\begin{equation} 
a_s[(b_{SM}-b_P)^2\,+\,b_{S}^2]=a_s\,b_{SM}^2\;, 
\end{equation} 
which leads to 
\begin{equation} 
(b_P-b_{SM})^2\,+\,b_{S}^2=b_{SM}^2\;, 
\label{b-cond}
\end{equation} 
or 
\begin{equation} 
R_S^2\,+\,\left[R_P-\frac{(m_b+m_s)}{m_{B_s}^2}b_{SM}\right]^2=
\left(\frac{m_b + m_s}{m_{B_s}^2} b_{SM} \right)^2 \;. 
\label{sm-cond} 
\end{equation} 
Eq.~(\ref{sm-cond}) represents a circle in $R_S-R_P$ plane  
with center at $\left(0,(m_b+m_s)b_{SM}/m_{B_s}^2 \right)$. 
 
The circle representing eq.~(\ref{sm-cond}) passes through
the origin ($R_S = R_P =0$), which corresponds to the SM.
However, in general the points on the circle have nonzero
$R_S$, and hence imply nonvanishing $A_{LP}$.
Therefore it is possible to have a nonzero value of  
$\alp$ even if $B(B_s\to \mu^+ \mu^-)$ is equal  
to its SM prediction. 
Thus $\alp$ can still serve as an important observable 
to search for SPNP even if $B(B_s\to \mu^+ \mu^-)$  
is observed to be very close to its SM prediction.

\subsubsection{LHCb fails to find $\bsll$}

If LHCb fails to find $\bsll$ or puts an upper bound on its 
branching ratio which is smaller than $2 \times 10^{-9}$
({\it $5\sigma$ below SM prediction}), this scenario can still be
accomodated within the SPNP.

The interference between the SPNP and SM operators
can decrease the branching ratio  
$B(B_s\to \mu^+ \mu^-)$ far below its SM prediction.  
In fact it can be seen from eq.~(\ref{blep_gen}), 
$B(B_s\to \mu^+ \mu^-)$ can even vanish, provided  
the following conditions are satisfied simultaneously:
\begin{equation} 
R_S=0,\,\,\,\, R_P=\frac{2m_\mu m_b}{m_{B_s}^2}R_A=0.04\,R_A\;. 
\end{equation} 
From Fig.~\ref{fig4}, it can be seen that for low
$R_S$ values, it is indeed possible to suppress 
$B(\bsll)$ much below its SM value.

\subsubsection{Both $B(\bsll)$ and $A_{LP}$ are consistent 
with the SM}  

The lepton polarization asymmetry is a result of the
interference of the scalar term with pseudoscalar / axial
vector, as can be seen from eq.~(\ref{eqn:alp}).
Therefore it vanishes when either $b_S$ or 
$(b_P-b_{SM})$, defined in eq.~(\ref{b-def}), vanishes. 
Thus there exists the interesting possibility of nontrivial 
SPNP even when both $B(\bsll)$ and $A_{LP}$ are consistent 
with the SM. 
This occurs when
\begin{eqnarray}
 b_S=0 \; , &  &  b_P = 2 \,  b_{SM} \; , \\
\quad {\rm or} \quad \quad b_S = b_{SM} \; ,  & & b_P = b_{SM} \; ,
\end{eqnarray}
as can be confirmed from eq.~(\ref{b-cond}).
Therefore, the absence of SPNP is not
guaranteed simply by the consistency of these
observables with the SM; more channels need to be
examined to rule out this possibility completely.

%%%%%%%%%%%%%%%%%%%%%%%%%%%%%%%%%%%%%%%%%%%%%%%%% 
\section{Conclusions} 
\label{concl} 

An order of magnitude enhancement in $B(B_s\to \mu^+ \mu^-)$  
is possible only due to SPNP operators. Apart from $B(B_s\to \mu^+ \mu^-)$, 
observables such as FB asymmetry of muons in $\bkll$ and LP asymmetry of muons in 
$B_s\to \mu^+ \mu^-$ are also sensitive to SPNP operators. In this paper we consider the 
constraints on possible SPNP contribution to these observables coming from  
the present upper bound on $B(B_s\to \mu^+ \mu^-)$.  
 
We find that $B(B_s\to \mu^+ \mu^-)$ 
puts very stringent constraint on SPNP contribution to  
$\afb$ and restricts its value to be less than $ \sim 1\%$. 
Such a small FB asymmetry is almost impossible 
to be measured in experiments. In the literature,  
$\left\langle A_{FB}\right\rangle$ 
of muons in $\bkll$ has been considered a promising measurement for 
probing SPNP operators. 
Our results show that the present upper bound on $B(\bsll)$ makes 
searching for SPNP through $\left\langle A_{FB}\right\rangle$ 
a futile exercise.   
 
On the other hand, the present upper bound on 
$B(B_s\to \mu^+ \mu^-)$ does not put any constraint on $\alp$.  
Indeed, $\alp$ can be $100\%$ even if $B(B_s\to \mu^+ \mu^-)$ is 
close to its SM prediction. 
$\alp$ is sensitive only to SPNP operators and hence its 
nonzero value will give direct evidence for a non-standard 
Higgs sector.

A simultaneous determination of $B(B_s \to \mu^+ \mu^-)$ and $\alp$  
will enable us to separate the new physics scalar and pseudoscalar 
contributions.  
Therefore it is worth considering this observable in experiments 
to probe the extended Higgs sector.

\begin{acknowledgments}
One of the authors, A. K. Alok would like to acknowledge 
Y. Grossman for his valuable suggestions.
\end{acknowledgments}

\end{document}